\documentclass[10pt,twocolumn,twoside]{optica-suppl-materials}
\setboolean{shortarticle}{false}

\usepackage{aas_macro}
\usepackage{graphicx}
\usepackage[textwidth=8em,textsize=small]{todonotes}
\usepackage{amsmath}
\usepackage{xspace}
\usepackage{url}
\usepackage{subfig}
\newcommand{\statsref}[2]{#1}

\usepackage{natbib}
\usepackage{bibentry}
\nobibliography*
\setcitestyle{numbers,sort&compress,square,super}

\title{Statistical methods in astronomy}

\author[1]{James P. Long}
\author[2,3,4]{Rafael S. de Souza}

\affil[1]{Texas A\&M University,
 College Station, TX,
 USA.}
\affil[2]{Department of Physics \& Astronomy, University of North Carolina at Chapel Hill, Chapel Hill, NC 27599-3255, USA.}
\affil[3]{MTA E\"otv\"os University, EIRSA ``Lendulet'' Astrophysics Research Group, Budapest, Hungary.}
\affil[4]{Instituto de Astronomia, Geof\'isica e Ci\^encias Atmosf\'ericas, USP, SP, Brazil.}

\affil[$\ast$]{E-mail: jlong@stat.tamu.edu}
\affil[$\star$]{E-mail: rafaeldesouza@alumni.usp.br}

\begin{abstract}
We present a review of data types and statistical methods often encountered in astronomy. The aim is to provide an introduction to statistical applications in astronomy for statisticians and computer scientists.  We highlight the complex, often hierarchical, nature of many astronomy inference problems and advocate for cross-disciplinary collaborations to address these challenges.
\end{abstract}

\begin{document}
\maketitle

\section{Introduction}

Astronomy has a \statsref{long history}{stat07961} of exploiting observational data  to estimate parameters and quantify uncertainty in physical models. Problems in astronomy propelled the development of many statistical techniques, from classical least squares estimation\cite{stigler1986history,hilbe2017} to  contemporary methods such as  nested sampling \cite{Skilling2004,Feroz2008}.

Late 20\textsuperscript{th} century advances in data collection, such as automation of telescopes and use of CCD cameras, resulted in a dramatic increase in data size and complexity, producing a surge in use and development of statistical methodology. Astronomers use these data sets for a diverse range of science goals, including modeling formation of galaxies, finding earth--like planets\cite{Foreman2014}, estimating the metric expansion of space, and classifying transients.
    
This article reviews common data types and statistical methodology currently in use  in astronomy, with the goal of making astronomical applications more accessible to methodological and applied statisticians.  A non-exhaustive selection of topics is covered in this article. We refer readers to the ``Further Reading'' section at the end of this text in which historical and methodological viewpoints of astrostatistics are presented.

In Section \ref{sec:data} we review three common types of astronomical data: images, spectra, and \statsref{time series}{stat07786.pub2}. In Section \ref{sec:methods} we discuss some statistical methods currently used in astronomy. Many of these methods are under active development within the statistics and computer science research communities. We conclude in Section \ref{sec:challenges} by describing one astrostatistics challenge, mapping the Milky Way halo with RR Lyrae stars, and the various statistical tools necessary for addressing this problem.

\section{Astronomical Data Types}
\label{sec:data}

\subsection{Image Data}

\begin{figure}
\begin{center}
\includegraphics[scale=.4]{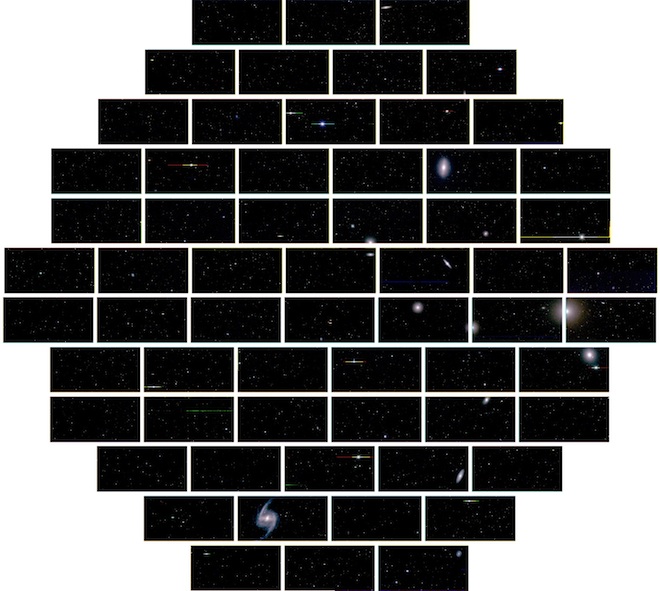}
\caption{Image of the night sky taken by the DECam. The white lines are gaps between the CCDs on the detector. Identifying, classifying, and estimating brightness of objects in images is a major statistical challenge in astronomy. \label{fig:des}}
\end{center}
\end{figure}

Telescopes take images of the night sky. Figure \ref{fig:des} shows an image taken by the Dark Energy Camera (DECam) as part of the Dark Energy Survey (DES).\footnote{\copyright AAS. Reproduced with permission. See \cite{flaugher2015dark} (doi:10.1088/0004-6256/150/5/150) for original publication.} DES takes approximately 400 one-gigabyte images  per night.\footnote{See \url{https://www.darkenergysurvey.org/the-des-project/survey-and-operations/data-management/}.} Astronomical images are often taken with a photometric filter which blocks certain light wavelengths.

A \textit{photometric pipeline} identifies objects in images and estimates their brightness. These pipelines contain many statistical tools such as machine learning algorithms (see Section \ref{sec:ml}) and hierarchical models (see Section \ref{sec:bayes}). The pipeline outputs a \textit{catalog} containing object positions, brightnesses, and classifications (star, galaxy, asteroid, etc.). Catalog data is typically much easier to study and model than the raw image data, so most subsequent analysis is performed on them.

\subsection{Spectral Data}

\begin{figure}
\includegraphics[width=\linewidth]{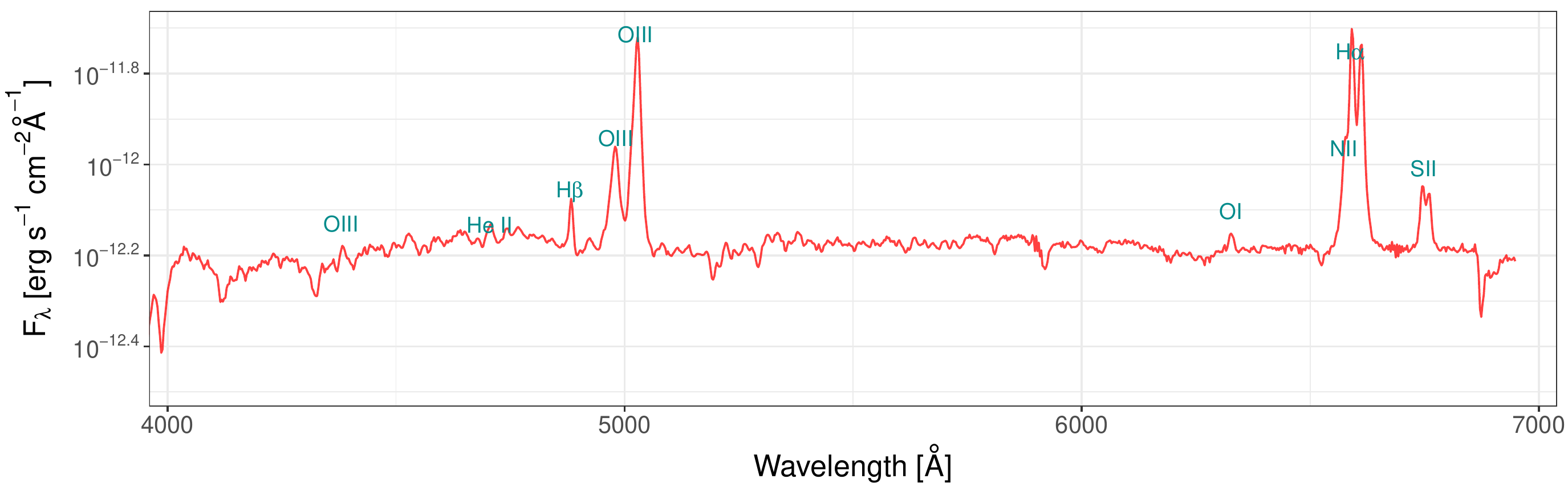}
\caption{Example of a galaxy spectra from Messier 77.}
\label{fig:galaxy_spectrum1}
\end{figure}

A spectra represents the intensity of light in different wavelengths, providing considerably more information than can be directly inferred from image data. Figure  \ref{fig:galaxy_spectrum1} shows the spectrum of the galaxy Messier 77, a barred spiral in the Cetus constellation. Spectra carry information about some of the most important physical properties of astronomical objects such as temperature and chemical composition. Additionally, the displacement of spectral features towards longer wavelength (known as redshift) may be used to estimate object distance, thus providing a precious tool for understanding evolution of the universe.

There are several astronomical surveys collecting spectral information such  as  the RAdial Velocity Experiment \cite{Kunder2017AJ},  one of the largest spectroscopic surveys of Milky Way stars publicly available. It enables study of Milky Way morphology and history through stellar spectroscopic observations and   astrometric databases. The SDSS-IV MaNGA Survey\cite{Bundy2015ApJ} collects $\sim$ 10,000 spectral measurements for nearby galaxies,  enabling construction of two-dimensional maps of physical properties throughout  each galaxy.

\subsection{Time Series and Functional Data}
\label{sec:timeseries}

Images and spectra represent two common forms of ``raw'' astronomical data, which together with photometric information, the integrated flux through a given filter, provide the basis to derive several data types. For example, many light sources vary in brightness as a function of time. Astronomical surveys which image the same area of the sky repeatedly over time produce a time series or \textit{light curve} for each object, permitting analysis of temporal brightness variation.

\begin{figure*}
\begin{center}
\centering
 \subfloat[Light Curve]{\label{fig:ogle}
 \includegraphics[scale=.4]{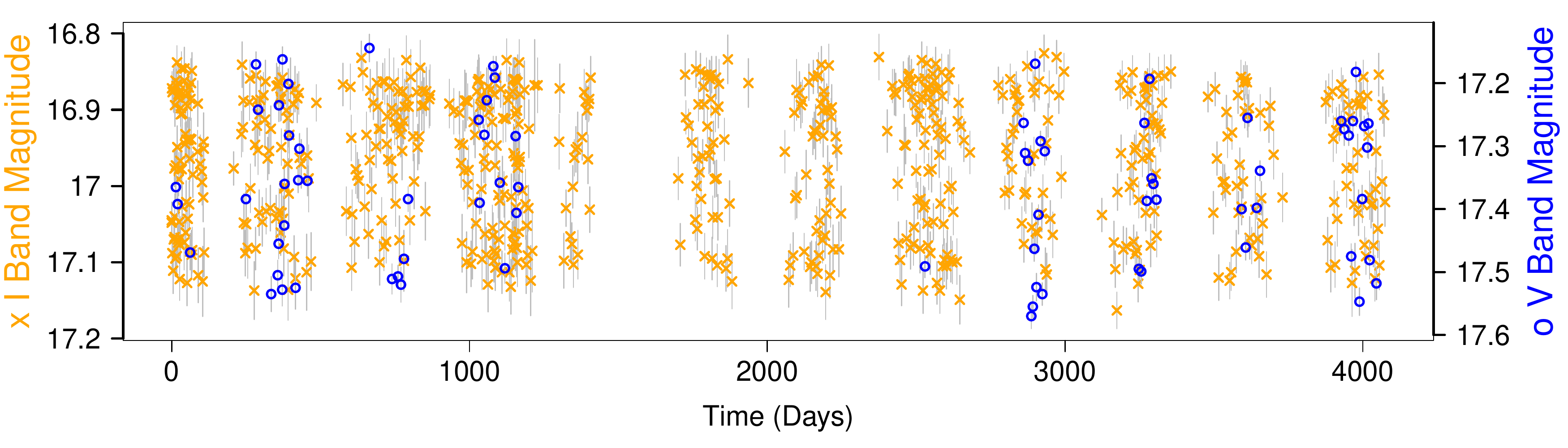}
 } \\
 \subfloat[Folded Light Curve]{\label{fig:ogle_folded}
 \includegraphics[scale=.4]{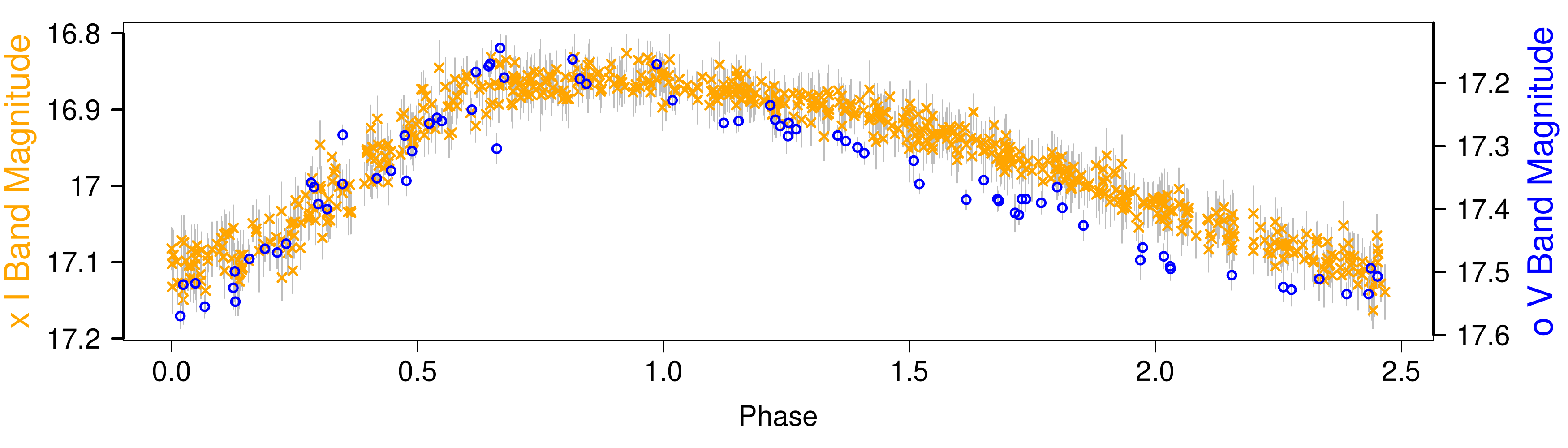}
 } \\
 \caption{(a) Light curve of a variable star observed by OGLE. Models from the time series and functional data analysis literature are often used for studying these objects. (b) The light curve in a) is produced by a periodic variable star. From the data in a) one can estimate a period ($\approx 2.48$ days) and plot the folded light curve, magnitude versus phase ($=$ time modulo period).}
\end{center}
\end{figure*}

\begin{figure}
\begin{center}
\includegraphics[width=0.85\linewidth]{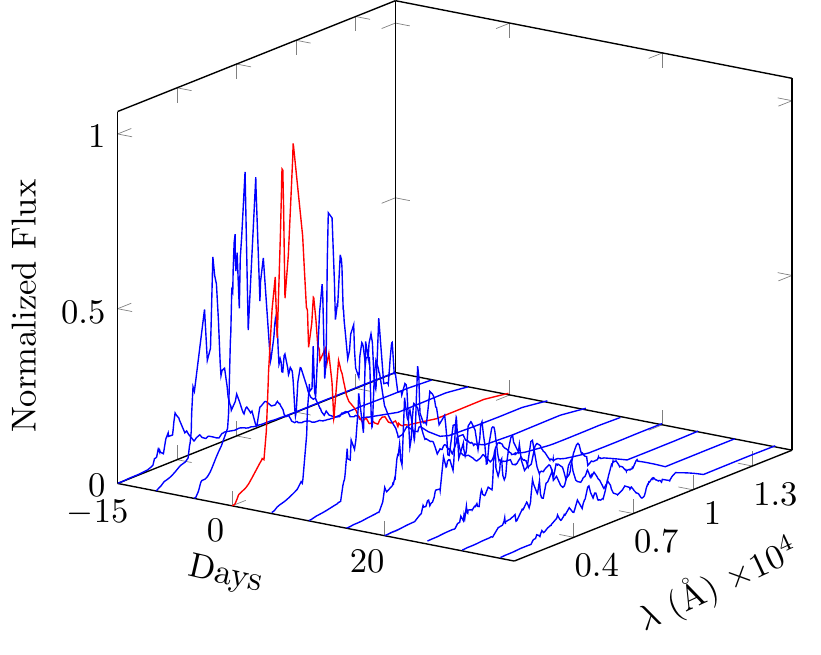}
\caption{Example of a supernova spectrum as function of days since maximum brightness.}
\label{fig:SNe_spectrum}
\end{center}
\end{figure}

Figure \ref{fig:ogle} shows a time series for a star observed by the Optical Gravitational Lensing Experiment (OGLE) \cite{udalski2008optical}. The data was collected in two filters, represented by orange crosses and blue circles over the course of approximately 10 years. The \textit{cadence}, or time spacing between observations, is irregular, a typical feature in astronomy data. OGLE has collected approximately 400,000 of these light curves, all of which are publicly available.\footnote{\url{http://ogledb.astrouw.edu.pl/~ogle/CVS/}} The statistical challenges with this data include modeling shape variation and classifying sources based on the astrophysical reason for brightness variation. For example, the star in Figure \ref{fig:ogle} is varying in brightness periodically over time. From this data, one can estimate a period and plot magnitude versus time modulo period (see Figure \ref{fig:ogle_folded}). Methods for estimating periods for this type of data are under active development\cite{long2016estimating,vanderplas2015periodograms}. For comparison, in  Figure \ref{fig:SNe_spectrum} we show a supernova spectra as a function of time with the epoch of maximum brightness highlighted in red.\cite{Kessler2010}

\section{Statistical Methodology in Astronomy}
\label{sec:methods}

Astrostatisticians use a wide range of statistical methods to analyze these complex data sets. We now discuss several areas of statistical methodology with recent applications within astronomy.

\subsection{Measurement Error Models}

Models typically assume homoskedasticity (i.e. errors with the same variance). Predictor (i.e. independent) variables in regression models are often assumed to be measured without error. However in Astronomy, \statsref{heteroskedastic errors}{stat03906} are the norm. Further, it is common practice to have estimates of the measurement error variances available through modeling of uncertainties inherent to the detection procedure\cite{feigelson2012modern}. In cases where the 
measurement error is large,  explicit \statsref{errors-in-variables models}{stat05747} are necessary to avoid  biased estimates, particularly in regression models. These models often have a hierarchical structure in which the true predictor values are treated as parameters.

Approaches commonly used in astronomy to solve this problem include the bivariate correlated errors and intrinsic scatter model \cite{Akritas1996} and  hierarchical Bayesian models \cite{kelly2007some,Loredo2004,hilbe2017,andreon2015bayesian}.  Examples of applications include  the development of a Gaussian mixture model for estimating a density from observations subject to measurement error\cite{bovy2011extreme}, the subsequent application of this model to account for flux uncertainties in probabilistic classification of quasars\cite{bovy2011think}, and use of a hierarchical Bayesian model to handle discrete measurement uncertainties in a negative binomial model to probe the population of globular clusters in galaxies\cite{deSouza2015MNRAS}. 

\subsection{Survival Analysis}

Astronomical surveys are,  by construction,  unable to obtain  unbiased samples from the population of objects. Surveys are often magnitude-limited, i.e. brighter objects are more likely to be detected.  This results in truncation due to the telescope sensitivity limit, which in astronomy is called \textit{Malmquist bias}\cite{Malmquist1922,Sandage2000}.  In other situations we know an object exists, but some of its features are too faint to be detected, resulting in censored observations.  In statistics, solutions for such problems are treated under the general umbrella of \statsref{survival analysis}{stat02177.pub2}.  Censoring and truncation are often called selection effects in astronomy\cite{feigelson2012modern}. Survival analysis challenges in astronomy may involve multivariate data\cite{schafer2007statistical}, nonparametric density estimation with truncation\cite{Efron1999}, or selection effects within a regression model\cite{kelly2007some}.

\subsection{Bayesian Models and Computation}
\label{sec:bayes}

Use of Bayesian methodology has grown considerably in astronomy over the past three decades. Active areas of Bayesian research include hierarchical models, posterior samplers, and models for complex data types such as images and functions.

\statsref{Hierarchical Bayesian models (HBM)}{stat07162} are used for problems where individual object parameters and population parameters are unknown. HBM are applied to many types of data in astronomy  as for instance to  detect and characterize galaxies in astronomical images using a HBM with variational inference to approximate the posterior\cite{regier2015celeste}, for modeling of supernovae light curves \cite{mandel2011type,sanders2015unsupervised}, and to fit cosmic ray data\cite{soiaporn2013multilevel}.

Approximate Bayesian computation (ABC) avoids computationally expensive likelihood evaluations by simulating data sets and comparing the distance between the simulated data and the actual data. ABC is being used in astronomy for inferring cosmological parameters\cite{schafer2012likelihood,weyant2013likelihood,Akeret2015,Jennings2017}  and  probing galaxy evolution \cite{Cameron2012,Hahn2017MNRAS}. The growing use of ABC has lead to the development of software packages, such as \textit{cosmoabc}, an ABC sampler via Population Monte Carlo for general astronomical applications\cite{ishida2015cosmoabc}. 

Several flavors and variants of \statsref{Markov Chain Monte Carlo}{stat07189} samplers have been developed by astronomers including an implementation of an affine--invariant ensemble sampler \cite{foreman2013emcee} and the Diffusive Nested Sampling, an extension of nested samplers \cite{brewer2011diffusive}.

\subsection{Generalized Linear Models}

The ubiquitous linear regression model relies on a number of distributional assumptions which fail to hold when the data come from \textit{exponential family} distributions other than the Gaussian.
\statsref{Generalized linear models (GLMs)}{stat05752}\cite{Neuhaus2011},  assume, through a link function, a linear relationship between the response variable $y$ and set of predictors $x$. Several problems in astronomy require the use of GLMs and extensions, such as modeling the  fraction  of Seyfert galaxies in  terms of environment (Bernoulli)\cite{deSouza2015glm,deSouza2016}, the population of globular clusters  as a function of the host galaxy properties (Negative binomial)\cite{deSouza2015MNRAS}, and  the  distance of galaxies  as a function of their  colors (Gamma)\cite{Elliott2015}.

\subsection{Machine Learning}
\label{sec:ml}

For several astronomy problems, prediction and pattern recognition are more important than parameter estimation. Methodology from the \statsref{machine learning (ML)}{stat05023.pub2} community is now routinely used to solve these problems \cite{Ball2010,Brescia2012}. While ``off the shelf" machine learning methods are sometimes sufficient, astronomy ML problems may involve additional challenges, such as biased training sets, computationally intensive feature extraction, or real--time classification, which inhibit use of standard methods. We describe some of these challenges below.

ML has seen extensive use in classification of variable source light curves (see Section \ref{sec:timeseries} for definition). Here astronomers are often more interested in determining the class of the source than estimating source parameters. Since light curves are functions, this is a functional classification problem. A common approach to the problem is to construct a training set of objects of known class (often using some level of human classification), extract features for these objects, and then train a ML classifier on this data. The classifier can then, in principle, be used to classify new objects in other surveys\cite{richards2011machine,Ishida2013,faraway2016modeling,Lochner2016}. ML tools are used for several other problems in astronomy including identification of sources in images \cite{brink2013using}, clustering of spectral data\cite{Sasdelli2016},  and photometric redshift estimation \cite{budavari2012photometric,freeman2009photometric}.  

A critical issue often overlooked is the lack of representativeness between  spectroscopic and photometric samples.  Cross-validation performance measures have been shown to be misleading in this situation.\cite{Beck2017} Mismatches between training and test samples are not exclusive to astronomical problems. Methodology developed by the ML community to address this challenge has been used on several astronomy problems, including domain adaptation \cite{Mahabal2017,Beck2017},   active learning \cite{Richards2012}, and a combination of both under the umbrella of  adaptive learning techniques \cite{Gupta2016}. However challenges remain, including incorporating feature measurement error, missing data, censoring and truncation into ML algorithms \cite{babu2015skysurveys,bloom2012automating}. 

\subsection{Information Visualization}

\begin{figure}
\includegraphics[width=\linewidth]{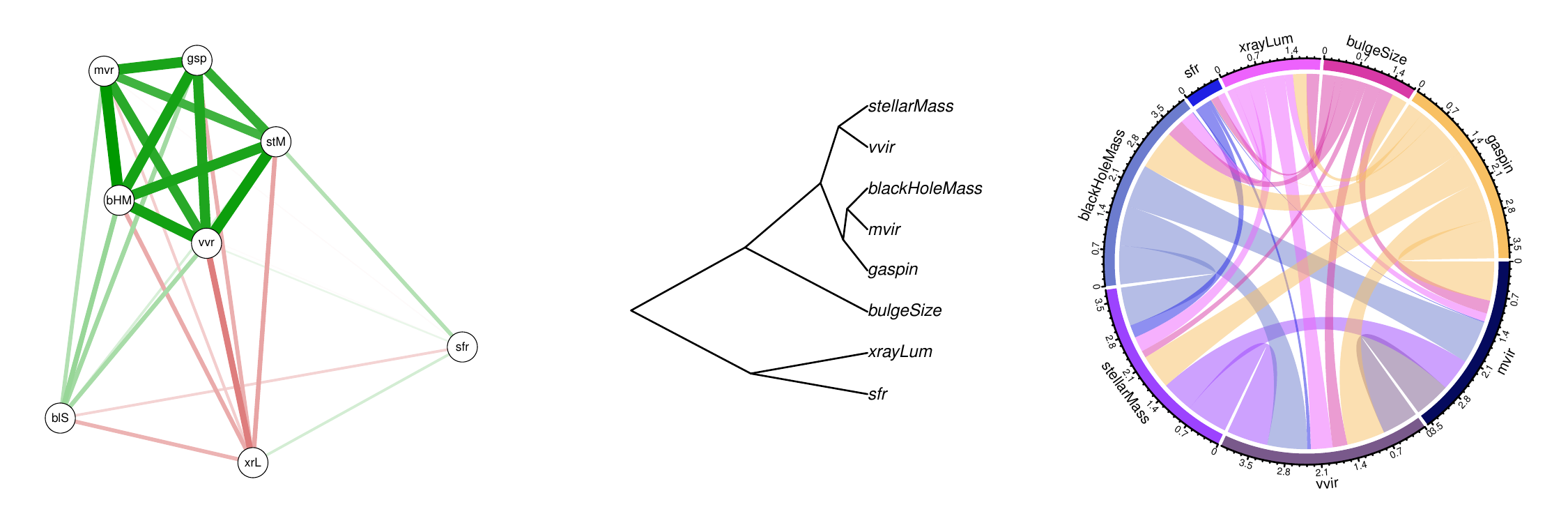}
\caption{Three visualizations (left to right: graph, cladogram and a chord diagram) of the same galaxy catalog from a N-body/hydro cosmological simulation.}
\label{fig:infoviz}
\end{figure}

\statsref{Visualization methods}{stat07365} exploit the human visual system to optimize intuitive insight into data structure. Whilst the role of visualization belongs to the groundwork of astronomical analysis, new paradigms for multidimensional data visualization are yet to be fully utilized. Patterns and non-trivial correlations that might go undetected in tabular-based data, can be unfolded if the proper tools are applied \cite{deSouza2015}. 
Among the methods that have been developed to facilitate the exploration of multivariate astronomical data are   phylogenetic trees \cite{Fraix2016}, graphs, chords  \cite{deSouza2015}, and starfish diagrams \cite{Konstantopoulos2015}. Figure \ref{fig:infoviz} shows a dataset from a N-body/Hydro cosmological simulation\cite{Guo2011}  visualized with three different techniques.
 
\section{Complex Inference Challenges in Astronomy: An Example}
\label{sec:challenges}

The process of turning data into scientific knowledge discovery typically requires the use of many statistical tools, often in innovative ways. Some of the most challenging statistical questions that arise in astronomy relate to how to merge these tools into a data analysis pipeline that permits valid statistical inferences while remaining computationally feasible.

As an illustrative example, consider the challenge of mapping the Milky Way halo, the region of space that surrounds our galaxy. This problem has attracted much recent attention.\cite{sesar2014stacking,sesar2016machine,sesar2009light,zinn2013silla,vivas2006quest} Astronomers would like to produce maps of the locations of stars in the halo and identify structures, such as collections of gravitationally bound stars. This has important consequences for the $\Lambda$ Cold Dark Matter ($\Lambda$CDM) cosmological model, our current framework for understanding how the universe was born and developed. Creating halo maps is difficult because it is impossible to determine the distance to most stars. We can, however, estimate distances to a small subset of stars, known as RR Lyrae (RRL), due to their all having similar luminosities (standard candles in astronomy). The locations of these stars trace the structure in the halo. Inference on the Milky Way halo requires:
\begin{enumerate}
\item Identifying the RRL stars among all stars observed in an astronomical survey. Recalling that variable stars are, as data, irregularly sampled functions (see Figure \ref{fig:ogle}), this is a large functional data classification problem. Once the RRL have been identified, we estimate their distance.
\item Using the estimated locations of the RRL, we estimate the local density of objects in order to identify structure. Often RRL locations are viewed as a realization from a Poisson process in three dimensional space. Errors from the previous step, including misclassified stars and uncertainty on distance estimates impact this map estimate.
\item Finally, one can compare the observed structure in the halo map to predictions made by the $\Lambda$CDM cosmological model. Different halo structures provide evidence for different values of the free parameters in $\Lambda$CDM. These comparisons could be heuristic or more quantitative (e.g. optimizing parameters in a cosmological simulation to produce halo structure which most closely resembles observations).
\end{enumerate} 

Schafer\cite{schafer2015framework} argues that cosmological inference problems are best divided into three stages: inference on object parameters, inferences on class parameters, and finally inferences on the fundamental cosmological parameters. The three steps above roughly correspond to these stages. Each stage requires many statistical decisions. Uncertainty must be propagated through the stages while at the same time approximations must be made to keep the analysis pipeline computationally feasible. 

Upcoming astronomical sky surveys, such as the Transiting Exoplanet Survey Satellite (TESS)\footnote{https://tess.gsfc.nasa.gov/}, the James Webb Space Telescope (JWST)\footnote{https://www.jwst.nasa.gov/}, and the Large Synoptic Survey Telescope (LSST)\footnote{https://www.lsst.org/}, promise ever larger data sets with more challenging inference problems. Interdisciplinary collaborations of statisticians and astronomers will be essential for developing the new statistical methodology necessary for fully realizing the science potential of these projects.

\section*{Related Articles}

\begin{itemize}
\item \bibentry{Feigelson04}
\item \bibentry{Goodman2012}
\item \bibentry{loredo2013bayesian}
\item \bibentry{stat07961}
\item \bibentry{SIGN:SIGN785}
\item \bibentry{Cameron2014IAUS}
\item \bibentry{schafer2015framework}
\item \bibentry{Feigelson2017IAUS}
\item \bibentry{Sharma2017}
\end{itemize}


\bibliographystyle{ieeetr}
\bibliography{ref}
\end{document}